\documentclass[11pt]{article}

\usepackage{amsmath}
\usepackage{amssymb}
\usepackage{caption}
\usepackage{subcaption}

\usepackage{graphicx}
\usepackage{authblk}
\usepackage[hypertexnames=false,colorlinks=true,linkcolor=blue,citecolor=blue,urlcolor=black]{hyperref}
\usepackage[numbers,comma,square,sort&compress]{natbib}
\usepackage[a4paper,text={6.5in,10in},centering]{geometry}

\usepackage[dvipsnames]{xcolor}
\newcommand{\rone}[1]{#1}
\newcommand{\rtwo}[1]{#1}
\newcommand{\new}[1]{#1}

\graphicspath{{eps/}{pdf/}}
\makeatletter\@addtoreset{equation}{section}\makeatother

\newcommand{\eps}[0]{\varepsilon}

 {\begin{trivlist} \item[]{\bf Proof. }}%
 {\hspace*{\fill}$\rule{.4\baselineskip}{.4\baselineskip}$\end{trivlist}}

 {\begin{trivlist}\item[]\textbf{Acknowledgments.}}{\end{trivlist}}
\newcommand{\bP}{\mathbf{P}}
\newcommand{\bA}{\mathbf{A}}

\newcommand{\bq}{{\Theta}}
\newcommand{\bx}{\mathbf{x}}

\begin{document}

\title{A coherent structure approach for parameter estimation in Lagrangian Data Assimilation}
\author[1]{John Maclean\thanks{corresponding author, available at \url{jmaclean@ad.unc.edu}}}
\author[2]{Naratip Santitissadeekorn}
\author[3]{Christopher KRT Jones}
\affil[1,3]{\small Department of Mathematics and RENCI, University of North Carolina at Chapel Hill}
\affil[2]{\small Department of Mathematics, University of Surrey, Guildford}

%\date{\today}
\maketitle
\begin{abstract}
We introduce a data assimilation method to estimate model parameters with observations of passive tracers by directly assimilating Lagrangian Coherent Structures. Our approach differs from the usual Lagrangian Data Assimilation approach, where parameters are estimated based on tracer trajectories. We employ the Approximate Bayesian Computation (ABC) framework to avoid computing the likelihood function of the coherent structure, which is usually unavailable. We solve the ABC by a Sequential Monte Carlo (SMC) method, and use Principal Component Analysis (PCA) to identify the coherent patterns from tracer trajectory data. Our new method shows remarkably improved results compared to the bootstrap particle filter \rtwo{when the physical model exhibits chaotic advection}.
\end{abstract}
%%%%%%%%%%%%%%%%%%%%%%%%%%%%%%%%%%%%%%%%%
%\begin{abstract}
%An approximating Kalman filter for the crime model is derived based on a (sequential) Bayesian inference. The ensemble-based implementation of the filter is employed in a similar manner to the Ensemble Kalman filtering (EnKF).
%\end{abstract}
\section{Introduction}
\label{Intro}
The problem of assimilating the path travelled by ocean instruments such as drifters to estimate states and/or parameters of dynamical systems is commonly known as Lagrangian data assimilation (LaDA). Every method for LaDA depends on the knowledge of error statistics for the observed position of the drifter paths; typically, it assumes the normal likelihood function and uncorrelated errors of observed position among all drifters. In this paper, we explore a novel idea for LaDA that assimilates a ``coherent (or persistent) structure" hidden in the Lagrangian path of the drifters instead of directly using the observed positions. We discuss the main advantage of this idea in the situation where the number of drifters is large, observed position has small variance and the flow is chaotic. We also demonstrate the improved accuracy of the parameter estimates of the new method when comparing with the particle filtering (PF) approach, which is apparent if the drifter paths are dominated by chaotic advection.
\par
Modern satellite-tracked ocean drifters~\cite{Dohan10,Argos} have been recognized as an important source of data for oceanographic and climate research. Drifters travel near the surface and measure physical data such as surface temperature along their paths via drifter sensors~\cite{Dimarco05}. In addition, they offer Lagrangian path data through the Doppler frequency shift on their satellite-based transmission. The accuracy of drifter locations can be less than 150 meters~\cite{Argos}, and there is a global ocean drifter set providing measurement data at a mean time interval of 1.2 hours spanning over a decade~\cite{Dohan10}.
\par
\rone{Oceanographic flows exhibit patterns at scales from metres to kilometres, and correspondingly drifters are seen to move in eddies or gyres and follow currents, creating small and large-scale patterns which can be used to infer the sea-surface flows \cite{Poulain01, Fratantoni01}. These patterns are associated with fundamental structures of the flow, including stable and unstable directions, centres and saddles. The goal of the current work is to infer the underlying flow state, or the parameter set governing the flow, \emph{from the patterns} rather than directly from the drifter trajectories.}
\par
Several LaDA methods have been presented in the last decade for sequentially assimilating Lagrangian path data into dynamical system models of the ocean~\cite{Ide02,Molcard03,Salman06,Spiller08,Apte08,LiuThesis,Krause09,Nara15a,Nara15b}. These methods append a Lagrangian model of the drifter locations to the Eulerian flow model in order to make inferences about how drifter locations correlate with model states; this approach also avoids the difficulty in which Lagrangian path data does not correspond to the fixed Eulerian grid locations. The key challenge of Lagrangian Data Assimilation (LaDA) is that the Lagrangian path of the passive tracers or drifters can be strongly nonlinear, which makes it difficult to sequentially estimate model state variables or parameters by computationally efficient methods such as the Ensemble Kalman filter (EnKF) \cite{Apte08}. It has been emphasized in several pieces of work that a nonlinear, Bayesian filtering framework such as the Particle Filter (PF) is more reliable in quantifying the uncertainty of the estimates~\cite{Apte08,Spiller08,LiuThesis,Apte13,Nara15a}. \rone{However, when the drifter locations are observed at a high accuracy (e.g. the class-three Argos), the observation is considered to be highly informative, corresponding to a likelihood probability with a small variance in the Bayesian framework. Assimilating highly-informative observations in a high-dimensional space is problematic since significant probability is contained in a region of extremely small ``volume''. The standard PF tends to perform very poorly in such a regime~\cite{Moral14}. With a narrow prior, it is probable that no particles will lie in a region of significant probability after a short amount of time. In this situation the PF weights, which fall off exponentially, assign probability one to a single particle and probability 0 to the rest.} Therefore, it would be difficult to apply PF or its variants to high-dimensional models of the ocean, or assimilate data from a large number of drifters, such as the approximately 2200 ocean drifters from the Global Ocean drifter Program (www.aoml.noaa.gov/envids/gld/). \rone{The problem of drawing inference from drifter trajectories is compounded by factors which tend to separate the numerically simulated drifters from the true trajectories. One such factor is model error, which is not considered in this paper. A second factor of particular importance to the current work is chaotic advection in the flow, which can readily create a situation where the simulated and observed drifter trajectories will appear unrelated, particularly if the observations are sparse.}\\  A hybrid PF-EnKF method has been proposed to reduce the computational cost in high-dimensional problems~\cite{Laura15, Poterjoy16}; in this approach it is assumed that the Eulerian flow model is high-dimensional and the Lagrangian model for the drifter locations is low-dimensional. A similar idea using a hybrid PF-EnKF to assimilate the Lagrangian path data has also been developed in~\cite{Nara15a,Nara15b} where EnKF is used to estimate drifter paths in tandem with PF for parameter estimation. However, to the best of our knowledge, the challenge of assimilating a high-dimensional drifter data set is still an open problem.
\par
%In all previous LaDA work, drifter locations and model parameters must be simultaneously estimated, which leads to a relatively high-dimensional problem just on account of the drifters if their number is high.
%We remark that the foregoing joint state-parameter estimation methods are excessive if our interest is only to estimate parameters, and not the drifter positions. 
These issues motivate us to exploit the qualitative structure of complex nonlinear flows - such as coherent structures or persistent patterns, if they exist - to make inferences on the model parameters.

%For example, if $N$ tracer locations are considered to be i.i.d. observations with a normal distribution, the likelihood probability, the product of $N$ independent, identical normal distribution with a small variance, will contain most of its probability in a very small volume in a high-dimensional space for a large $N$. The computational cost for such a problem can be prohibitively expensive. These issues motivate us to exploit the qualitative structure of complex nonlinear flow to make inferences on the model parameters, rather than directly assimilating the Lagrangian path data.
\par
In particular, we will exploit a large-scale ``Lagrangian coherent pattern" or just ``coherent pattern" hidden in the Lagrangian path data. The coherent pattern is imprecisely defined as a region in state space, for example a coherent vortex or nonlinear jet, that moves along with the flow without dispersing. Coherent patterns must also be robust under small diffusive peturbations, that is, the coherent object should still hold its geometric structure together under some diffusion process. When the flow is known either numerically or in a closed form, these coherent regions can be identified by several approaches. In the probabilistic approach, the almost invariant-set framework was developed for autonomous and periodic flows by using the transfer (or Perron-Frobenius) operator~\cite{Dellnitz97,Dellnitz99,Billing02,Froyland03} or the infinitesimal generator~\cite{Froyland13fast}; the latter reduces the high computational cost of the former approach. For nonautonomous flows, the finite-time coherent set method via a transfer operator was introduced for the first time in~\cite{Nara10a,Nara10b,Nara10c} and it was applied to detect long-lived vortices such as the stratospheric polar vortex~\cite{Nara10b,Nara10c} and Agulhas rings~\cite{Nara12}, both as two and three dimensional coherent objects. This method has a strong connection with spectral clustering as described in Chapter of 4 of~\cite{Narabook}. The coherent region as used in these works probabilistically minimizes the mass transport in and out of the region with respect to a reference measure (not necessarily the invariant measure), see~\cite{Nara10b,FroylandPadberg09} for full details.
\par
Unfortunately, all of the above coherent set identification methods require a set of governing equations of the underlying dynamics and cannot be applied to identify the coherent pattern directly from the Lagrangian path data, which is an essential goal of using LaDA in the current work. Recently, new data-driven algorithms have been developed that can extract coherent patterns from possibly sparse Lagrangian path data without having to rely on governing equations. To name a few, these methods include the diffusion-map algorithm~\cite{Banisch16}, fuzzy c-mean algoritm\cite{Froyland15rough}, bipartite spectral clustering~\cite{Hadjighasem_etal} and dynamic mode decomposition (DMD)~\cite{Schmid10,Tu14}. These methods can capture the time-dependent coherent pattern that represents a slowly-decaying mode of a complex flow field.
\\ 
It is not a primary concern of this paper to %study the spatio-temporal coherent patterns in a general context,
 select or consider the ideal algorithms to identify the spatio-temporal coherent patterns in a general context, but to advocate for a methodology in which identified coherent patterns are assimilated %in order to estimate model parameters
 in a Bayesian framework. For this reason, numerical examples in this paper will be limited to the case of a stationary spatial pattern where an application of standard principal component analysis (PCA) is satisfactory for the coherent pattern identification. We will provide a brief review of PCA in Section~\ref{sec:PCA}.
\par
In Bayesian data assimilation, a closed form of the likelihood function for the coherent pattern must be known. However, even if the likelihood of tracer positions is commonly known, it is still a difficult task to derive the likelihood function of any corresponding coherent patterns. We will address this issue by applying the so-called ``Approximate" Bayesian computation (ABC) \cite{MarinEtAl12}, which can be thought of as a ``likelihood-free" Monte Carlo method. Originally, the ABC method was proposed to avoid evaluating a likelihood function that is computationally expensive by constructing a ``distance function" together with a summary statistic in a rejection-acceptance algorithm. In our context, this summary statistic is the coherent pattern. It is typically impossible to construct the required distance function without knowledge of the likelihood function of the coherent pattern; we will argue that given the likelihood function is unknown, a distance function with minima in the most likely regions of parameter values is a useful surrogate. We will discuss our choice of the distance function in Section~\ref{sec:KL} and ABC methods in Section~\ref{sec:ABC}.\\

\section{Coherent spatial patterns via PCA}\label{sec:PCA}
As we will make inference about model parameters based on the coherent pattern of the flow, we use the dominant eigenvector (corresponding to the eigenvalue with the largest magnitude) of the covariance matrix as a representative of the coherent pattern, as conventionally performed by the Principal Component Analysis (PCA)~\cite{Jolliffe}. Note that we focus on assimilating only the most ``dominant" pattern in this work, though several patterns may exist for the underlying flow. A given trajectory of $d$ tracers, consisting of recordings of the $x_1$ and $x_2$ coordinates at times $t_0<t_1<t_2<\cdots<t_n$, is denoted by $\bx_{0:n}:=\{\bx_{0:n}^{(i)}\}_{i=1}^d$, where $\bx_{0:n}^{(i)}=\{(x_{1}^{(i)}(t_0),x_{2}^{(i)}(t_0)),\ldots,(x_{1}^{(i)}(t_n),x_{2}^{(i)}(t_n))\}$. Let $x_{1,0:n}^{(i)}=[x_{1}^{(i)}(t_0),\ldots,x_{1}^{(i)}(t_n)]$ for $i=1,\ldots,d$ and $\tilde{\bA}_{x_1}$ be a $d\times n$ matrix whose $i-$th row is $x_{1,0:n}^{(i)}$.
We define the anomaly matrix of the $x_1$ coordinate by $\bA_{x_1}=\tilde{\bA}_{x_1}-\bar{\bx}_1\otimes\mathbf{1}_d$, where the vector $\bar{\bx}_1$ denotes the average of the tracer position over their trajectory and $\mathbf{1}_d$ is a $d$-dimensional vector with all entries 1. The covariance matrix can then be constructed by
\begin{equation}\label{eq:Px}
\bP_{x_1} = \frac{1}{d-1}(\bA_{x_1}\bA'_{x_1}).
\end{equation}

The dominant eigenvector of $\bP_{x_1}$ gives an approximation of the coherent pattern according to the correlation structure of the $x_1$-coordinate of the tracer trajectories. In general, one may have to consider different patterns that may be captured by eigenvectors with smaller eigenvalues. The coherent pattern described by the correlation structure in the $x_2-$direction can be similarly extracted from $\bP_{x_2}$, which can be constructed in a similar manner. For the numerical experiment in this work, it will be adequate to use only $\bP_{x_1}$ to identify coherent patterns. In more general cases, recent advanced techniques~\cite{Banisch16,Froyland15rough,Hadjighasem_etal} to identify coherent patterns from the data may be required. We remark that the time scale is important to observe the coherent patterns. Thus, $t_n$ has to be ``long enough", which will depend on the underlying dynamics.

%%%%%%%%%%%%%%%%%%%%%%%%%%%%%%%%%%%%%%%%%%%%%%%%%%%%%%%%%%%%%%%%%%%%%%%%%%%%%%%%
\section{Hellinger distance}\label{sec:KL}
In order to implement ABC, we need a distance function. We adopt the Hellinger distance to model the ``distance" between the predicted spatial pattern $f\in\mathbb{R}^d$, and the observed pattern $g\in\mathbb{R}^d$, both of which are identified by performing PCA on the tracer trajectory data. We normalize $f$ and $g$ so that their entries are between 0 and 1 and summed to unity. The Hellinger distance, see~\cite{KodyBook}, between $f$ and $g$ is given by
\begin{equation}\label{eq:Hellinger}
    H(f,g)=\frac{1}{\sqrt{2}}\|\sqrt{f}-\sqrt{g}\|_2.
\end{equation}
Therefore, $H(f,g)$ satisfies the triangle inequality and $0\leq H(f,g)\leq1$ for all $f$ and $g$ normalized as above. When comparing the patterns, the difference in the observed positions of the tracers and the predicted positions is irrelevant. What is important is that we are able to group together the tracers with similar dynamical fates; for example, tracers in a persistent vortex should be approximately clustered as a group, separating them out from the tracers in a strongly mixing or turbulent region. The preceding remark is especially crucial in assimilating a large number of tracers for noisy dynamical systems or chaotic flows because the model prediction of tracer positions can be inaccurate due to the noisy peturbation in the flow, while the coherent pattern is typically robust under peturbation~\cite{Haller02}.

%In order to use this distance in the context of ABC, we need to transform the likelihood in observations (i.e. the observation errors for the tracers) to the likelihood of the %$KL(f,g)$. We assume that the observation error statistic is known and we can use it to approximate the error statistics for $KL$. Therefore, our ``observation space" now becomes %only one-dimensional! Statistically speaking, the $KL$ will serve as a summary statistics to make an inference about the the unknown parameters.
%\par
%Remarks:
%\begin{itemize}
%  \item When comparing the pattern, the locations of tracers are not important. What is important is that we are able to ``label" the tracer so that we can compare the ``PCA score" between the observed and simulated tracers, which, in general, are not at the same position.
%  \item It may not matter so much if we were to use another metric instead of the KL divergence. Perhaps, $L_1$ or $L_2$ norm may also work.
%  \item We could also use $$KL(f,g)+KL(g,f)$$ if a symmetric distance is really required.
%\end{itemize}

%%%%%%%%%%%%%%%%%%%%%%%%%%%%%%%%%%%%%%%%%%%%%%%%%%%%%%%%%%%%%%%%%%%%%%%%%%%%%%%%
\section{Likelihood function of coherent structures}
Although considerable research has been devoted to the augmented state approach for LaDA (where model parameters are augmented to the state), rather less attention has been paid to the situation where the tracer positions are not of interest as their positions are observed quite accurately; in this situation the tracer positions are ``nuisance" random variables in the inference problem.
\par
Suppose that the random variable $\theta$ aggregates all unknown parameters, and a random variable $y$ represents the coherent pattern identified by the principle component. At the heart of our proposed LaDA method for parameter estimation is the inference problem
\begin{equation}\label{eq:maininference}
\pi(\theta|y)\propto f(y|\theta)\pi(\theta),
\end{equation}
where $f(y|\theta)$ is the likelihood function and $\pi(\theta)$ is the prior distribution.  Due to the uncertainty of the tracer locations $\bx_{0:n}$, the likelihood function of $y$ conditional on $\theta$ is obtained by
\begin{equation}\label{eq:nuisance}
f(y|\theta) = \int\tilde{f}(y|\theta,\bx_{0:n})\pi(\bx_{0:n}|\theta) d\bx_{0:n},
\end{equation}
where $\tilde{f}(y|\theta,\bx_{0:n})$ is the likelihood function of $y$ conditional on $\theta$ and $\bx_{0:n}$. The likelihood function $\tilde{f}(y|\theta,\bx_{0:n})$ is usually intractable due to a lack of knowledge of its analytic form, and so is \eqref{eq:nuisance}. One way to simplify the problem is to substitute the random variable $\bx_{0:n}$ in~\eqref{eq:nuisance} with some reference value $\hat{\bx}_{0:n}(\theta)$ (e.g. the mean, a sample, etc.). By doing so, we have
\begin{equation}\label{eq:nuisanceapprox}
f(y|\theta)\approx\tilde{f}(y|\hat{\bx}_{0:n}(\theta))\equiv\tilde{f}(y|\theta,\hat{\bx}_{0:n}).
\end{equation}
As a consequence, the inference problem~\eqref{eq:maininference} can be approximated by
\begin{equation}\label{eq:maininferenceapprox}
\pi(\theta|y)\propto\tilde{f}(y|\hat{\bx}_{0:n}(\theta))\pi(\theta).
\end{equation}
We emphasize that the approximated likelihood $f(y|\theta)$ in\eqref{eq:nuisanceapprox} still lacks an analytic form; hence, the evaluation of $f(y|\theta=\theta^\ast)$ for a given parameter value $\theta^\ast$ is still unavailable. We can, however, simulate the coherent pattern $y$ for $\theta^\ast$, which allows us to find the Hellinger distance between the predicted pattern (extracted from a sample of $\bx_{0:n}$) and the observed pattern $y$ (extracted from the observed tracer trajectories).
\par
Although the statistical approximation properties of~\eqref{eq:maininferenceapprox} are not examined herein, we heuristically argue that the approximation~\eqref{eq:nuisanceapprox} is reasonable in the current application. As pointed out earlier, the coherent pattern is robust to stochastic peturbation of the velocity field, so most samples of Lagrangian path $\bx_{0:n}$ would remain within the same flow regime. For example, even under the stochastic peturbation, most samples of the tracers in the recirculation regions tend to be in the recirculation region and those in the jet stream would remain in the jet stream as long as they are propagated under the same model parameters. It is the model parameters that prominently determine the large-scale coherent patterns of the flow. Thus, the likelihood $\tilde{f}(y|\theta=\theta^\ast,\bx_{0:n})$ for a fixed value $\theta^\ast$ would have almost the same value for any sample of $\bx_{0:n}$, so the approximation~\eqref{eq:nuisanceapprox} should be acceptable. In the next section, we describe a framework to solve the Bayesian computation of~\eqref{eq:maininferenceapprox} that only uses a distance function to avoid the difficulty of evaluating the likelihood function.

 %This is relevant to the LaDA since the Lagrangian path data, represented by $\bx$, under a noisy or chaotic flow would be very different even if they have a small difference in the initial condition. However, as pointed out earlier, the coherent pattern is robust to small peturbation of the velocity field, so it is the model parameters that prominently determine the large-scale coherent pattern of the flow; hence, the error in Lagrangian path itself would not play a significant role in the identification of the pattern as long as most of the tracers always stay within the same flow regime. If the model is reasonably accurate (up to a small stochastic peturbation), the approximated inference problem~\eqref{eq:maininferenceapprox} can be useful.

%%%%%%%%%%%%%%%%%%%%%%%%%%%%%%%%%%%%%%%%%%%%%%%%%%%%%%%%%%%%%%%%%%%%%%%%%%%%%%%%
\section{Approximate Bayesian computation (ABC)}\label{sec:ABC}
Given a prior density $\pi(\theta)$ of the parameter $\theta\in\Theta$, consider a situation where a likelihood function $f(y|\theta)$ of observations $y\in\mathcal{D}$ is not available in a closed form or its evaluation is computationally intractable. It is then difficult to use standard Bayesian computation methods to sample the posterior density $\pi(\theta|y)$.
Approximate Bayesian computation (ABC) is a likelihood-free monte carlo approach that only requires a simulation (or sampling) from the likelihood $f(y|\theta)$ but avoids an evaluation of $f(y|\theta)$. The foundation of ABC algorithms was suggested by Rubin as early as 1984~\cite{Rubin84}: that if we observe a discrete random variable $y^o\sim f(y|\theta)$, then we can sample $\pi(\theta|y)$ by jointly sampling $\theta^\ast\sim\pi(\theta)$ and $y^\ast\sim f(y|\theta^\ast)$, accepting $\theta^\ast$ and the pseudo observation $y^\ast$ only if $y^\ast=y^o$. The accepted values of $\theta^\ast$ will distribute according to the posterior density.
\par
In the ABC framework, the acceptance-rejection step replaces the impractical condition $y^\ast=y^o$ with $\rho(\eta(y^\ast),\eta(y^o))<\eps$, where $\rho$ is some measure of discrepancy, $\eps$ is a threshold level, and $\eta$ is some summary statistic. The threshold $\eps$ controls the trade-off between computational cost and accuracy; for a small $\eps$, the sample will give a good representation of the true posterior distribution but require more computation. Thus, all ABC algorithms draw the sample from the approximate posterior density
\begin{equation}
\label{ABC.f}\pi_\eps(\theta,y|y^o)\propto f(y|\theta)\pi(\theta)I_{\eps}(y),
\end{equation}
where 
\begin{align}\label{Ieps}
I_{\eps}(y)=
\begin{cases}
1\qquad\text{for } \rho(\eta(y),\eta(y^o))<\eps,\\
0\qquad\text{otherwise}.
\end{cases}
\end{align}
Provided that $\eps$ is sufficiently small and $\eta$ is a sufficient statistic, we hope that
\begin{equation}
\pi_\eps(\theta|y^o) = \int\pi_\eps(\theta,y|y^o)dy\approx\pi(\theta|y^o)
\end{equation}
\par
The simplest ABC algorithm draws a sample of $\theta$ by the following acceptance-rejection algorithm~\cite{Tavare97,Sisson07}:
\begin{description}
  \item[Algorithm 0]
  \item[Step 1.] Sample $\theta\sim\pi(\theta)$
  \item[Step 2.] Sample $y\sim f(y|\theta)$
  \item[Step 3.] Accept $\theta$ if $\rho(\eta(y),\eta(y^o))\leq\eps$
\end{description}

In the current application, $\theta$ is the model parameters, $y^o$ is the observed dominant spatial pattern, $y$ is the predicted spatial pattern, $\rho(y,y^o)$ is a distance function designed by the user, which is the Hellinger distance in this work, and $\eta$ is the identity function (i.e. $\eta(y)=y$). In the second step of the algorithm, we use the PCA method for pattern identification. Unfortunately, the above algorithm is ineffective due to a low acceptance rate, especially in the case where the prior and posterior distributions are very different. To improve the algorithm efficiency, several algorithms have been recently proposed, e.g., the adaptive SMC-ABC~\cite{Beaumont09, Tina08, MoralABC11}, which incorporates a sequential Monte Carlo (SMC) sampler~\cite{MoralSMC06} to approximate the posterior density by a collection of weighted ``particles" $\{\theta^{(i)},w^{(i)}\}$  for $i=1,\ldots,N$.
We will only outline the SMC-ABC algorithm used in our work and refer the details and discussion of the SMC-ABC in general to~\cite{MoralSMC06, MoralABC11}.

%%%%%%%%%%%%%%%%%%%%%%%%%%%%%%%%%%%%%%%%%%%%%%%%%%%%%%%%%%%%%%%%%%%%%%%%%%%%%%%%
\subsection{SMC-ABC}
Suppose that we wish to approximate a probability density $\pi(z)$ by random samples, where in the current context $z=(\theta,y)$, the collection of parameter estimates together with the corresponding predicted coherent patterns.
The SMC sampler is an approach to approximate a sequence of probability distributions $\tilde{\pi}_n(z_{0:n})$ for $n=1,\ldots,T$ and $z_{0:n}\equiv(z_0,\ldots,z_n)$ that admits $\pi_n(z_n)$ as a marginal distribution and satisfies $\pi_T(z_T)\approx\pi(z)$. Del Moral et al.~\cite{MoralABC11} proposed the following sequence
\begin{equation}\label{eq:SMAABC}
\tilde{\pi}_n(z_{0:n}) = \pi_n(z_n)\prod_{j=0}^{n-1}L_j(z_{j+1},z_j),
\end{equation}
where $L_n(z_{n+1},z_n)$ is the probability density of moving a given $z_{n+1}$ backward to $z_n$. Thus, given weighted samples $\{z_{n-1}^{(i)},w_{n-1}^{(i)}\}$ at the time step $n-1$, the SMC sampler draws $z_n^{(i)}\sim K_n(z_{n-1}^{(i)},\cdot)$ via the (forward) Markov transition kernel $K_n$ and updates the weight according to
\begin{equation}
w_n^{(i)}\propto w_{n-1}^{(i)}\frac{\pi_n(z_n^{(i)})L_{n-1}(z_n^{(i)},z_{n-1}^{(i)})}{\pi_{n-1}(z_{n-1}^{(i)})K_{n}(z_{n-1}^{(i)},z_{n}^{(i)})}.
\end{equation}
The above weights are then normalized so that $\sum_{i=1}^Nw_n^{(i)}=1$. A resampling step may be required to prevent the so-called particle collapse in which most particles have negligible weights. The degree of particle collapse is typically monitored by the effective sample size (ESS) given by
\begin{equation}
\label{ESS} N_{eff}\left(\{w_n^{(i)})\}\right):= \biggl(\sum_{i=1}^N(w_n^{(i)})^2\biggr)^{-1}.
\end{equation}
The resampling step is triggered when $N_{eff}<\delta N$ for some $0<\delta<1$.
\par
Again, we consider $z=(\theta,y)$ and the target distribution is $\pi_\epsilon(\theta,y|y^o)$. We are interested in sampling a sequence $\pi_{\epsilon_n}(\theta,y|y^o)$ for $\epsilon_0>\epsilon_1>\cdots>\epsilon_T=\epsilon$ and the required marginal $\pi_\epsilon(\theta|y^o)$ is obtained by the construction in~\eqref{eq:SMAABC}. The optimal choice of the backward kernel $L_n$ is given in~\cite{MoralABC11} but it is typically difficult to implement. A more convenient choice of $L_n$ is an MCMC kernal of invariant distribution $\pi_n$ associated with $K_n$:
\begin{equation}
L_{n-1}(z,z') = \frac{\pi_n(z')K_n(z',z)}{\pi_n(z)}.
\end{equation}
This leads to a weight update equation
\begin{equation}
\tilde{w}_n^{(i)}\propto\frac{\pi_{\epsilon_n}(\theta^{(i)}_{n-1},y^{(i)}_{n-1}|y^o)}{\pi_{\epsilon_{n-1}}(\theta^{(i)}_{n-1},y^{(i)}_{n-1}|y^o)}w_{n-1}^{(i)}
\end{equation}
According to~\eqref{ABC.f}, the weight update equation is given by
\begin{equation}\label{eq:weightupdate}
\tilde{w}_n^{(i)}=
\begin{cases}
w_{n-1}^{(i)}{I_{\epsilon_n}(y^{(i)}_{n-1})}\qquad\text{for } w_{n-1}^{(i)}>0\\
0\qquad\qquad\qquad\qquad\!\text{otherwise}
\end{cases},
\end{equation}
where $I_\epsilon$ is defined in \eqref{Ieps}. The pseudo-weights $\tilde{w}_n^{(i)}$ can be normalized, yielding $w_n^{(i)}=\tilde{w}_n^{(i)}\left(\sum_{i=1}^N\tilde{w}_n^{(i)}\right)^{-1}.$  It is also possible to use repeated simulations of the pseudo observations for a given parameter, rather than using a single pseudo observation as done here, to reduce the variance of the particle weights. This generalization is discussed in~\cite{MoralABC11}.
\par
We also follow~\cite{MoralABC11} to design an adaptive schedule for the tolerance levels, $\epsilon_0>\epsilon_1>\cdots>\epsilon_T=\epsilon$. In particular, $\epsilon_n$ is chosen to satisfy the criterion
\begin{equation}\label{eq:newepsilon}
\#\{w_n^{(i)}:w_n^{(i)}>0 \} = \alpha\,\#\{w_{n-1}^{(i)}:w_{n-1}^{(i)}>0\} \;,
\end{equation}
for $\alpha\in(0,1)$, \rtwo{and where $\#\{.\}$ denotes the number of elements in the set $\{.\}$}. \rtwo{In practice the preceding three equations are solved in reverse order. One sorts the particles and associated trajectories $(\theta_{n-1}^{(i)},y_{n-1}^{(i)})$ according to their (Hellinger) distance $\rho(y_{n-1}^{(i)},y^o)$. Of the particles with non-zero weight, a proportion $1-\alpha$ with the largest `distance' as measured by $\rho$ has weight set to 0. The resulting weights are the desired weights $w_n^{(i)}$ for the $n$-th step; the tolerance level $\epsilon_n$ is given by the largest Hellinger distance among the non-zero weight particles, $\max_{\,n} \left\{ \rho(y_{n}^{(i)},y^o) | w_n^{(i)}>0\right\} $.  }\\
We implement the the MCMC kernel $K_n(z,z')$ by the Metropolis-Hastings algorithm. In particular, we draw a candidate $z^\ast=(\theta^\ast,y^\ast)$ from a proposal $q_n(\theta,\theta^\ast)f(y^\ast|\theta^\ast)$, where $q_n(\theta,\theta^\ast)$ is a transition kernel for $\theta$. It is then accepted with probability given by the MH ratio
\begin{equation}\label{eq:MHratio}
1\wedge\frac{I_{\epsilon_n}(y^\ast)}{I_{\epsilon_{n-1}}(y)}\frac{q_n(\theta^\ast,\theta)}{q_n(\theta,\theta^\ast)}\frac{\pi(\theta^\ast)}{\pi(\theta)}.
\end{equation}
\par
Note that the above MH acceptance-rejection process is undertaken only for the particles with non-zero weights since the zero-weighted particles will keep their zero weights until being ``resurrected" via the Resampling algorithm.\\

\rtwo{The weight update equations~\eqref{eq:newepsilon}~and~\eqref{eq:weightupdate} ensure that the maximal distance between the pattern associated with any particle and the observed pattern, as measured by \eqref{Ieps}, decreases. One immediate consequence of this is that one cannot add noise to the particles when resampling. If noise is added, some or indeed most of the refreshed particles will have associated distances larger than the threshold $\epsilon_n$, and the monotonic convergence of the algorithm will be violated. Rather than refreshing the particles by resampling, the Metropolis-Hastings algorithm in steps 3-4 serve to continually refresh the available particles. In particular the accept-reject criterion for the algorithm depends on the current threshold $\epsilon_n$ and grows more stringent as time passes.}\\
\par
The SMC-ABC algorithm is summarized as follows.
\begin{description}\label{algo_main}
  \item[Algorithm 1]
  \item[Initialization] Set $\epsilon_0=\infty$, $w_0^{(i)}=1/N$ for $i=1,\ldots,N$ and $n=1$. Draw $\theta_0^{(i)}\sim\pi(\cdot)$ and $y^{(i)}\sim f(\cdot|\theta_0^{(i)})$.
  \item[While] $\epsilon_{n-1}>\epsilon$
  \begin{description}
    \item[step 1] Solve~\eqref{eq:newepsilon}~and~\eqref{eq:weightupdate} jointly for $\epsilon_n$ and $w_n^{(i)}$.
    %\item[step 2] Update $w_n^{(i)}$ according to~\eqref{eq:weightupdate}.
    \item[step 2] If $N_{eff}(\{w_n^{(i)})\})<N_T$, do Resampling.
    \item[for] $i=1,\dots,N$
    \begin{description}
    \item[step 3] $(\theta^\ast,y^\ast)\sim q_n(\theta_n^{(i)},\theta^\ast)f(y^\ast|\theta^\ast)$.
    \item[step 4] Accept $\theta_{n+1}^{(i)} = \theta^\ast$ with the probability~\eqref{eq:MHratio}; otherwise $\theta_{n+1}^{(i)} =\theta_{n}^{(i)}$.
    \end{description}
    \item[end for]
    \item[step 5] Set $n=n+1$.
  \end{description}
  \item[End while loop]
\end{description}

In order to assess the viability of PCA as a tool in data assimilation, we will compare the above SMC-ABC algorithm to several other schemes. In particular, we will consider the standard particle filter (see for instance \cite{Moral96,Wikle07}). We will estimate the likelihood of the observations in the particle filter by assuming that their error statistics are distributed normally. \\%Our intent is to demonstrate numerically the shortcomings of the particle filter claimed in Section~\ref{Intro}, and to show the competitive advantages of PCA. \\% in assimilating many observations in a noisy dynamical system.\\
We outline a particle filter in the next section.

%%%%%%%%%%%%%%%%%%%%%%%%%%%%%%%%%%%%%%%%%%%%%%%%%%%%%%%%%%%%%%%%%%%%%%%%%%%%%%%%
\section{Particle Filters}
For comparison with SMC-ABC that is used to assimilate coherent patterns, we use the standard (bootstrap) particle filter (PF)~\cite{GordonSalmond93}, where the importance density is the same as the transition density, to assimilate the tracer trajectories in the standard manner done in previous research~\cite{,Spiller08,Apte13,LiuThesis,Nara15a,Nara15b}. It should be emphasized that while SMC-ABC estimates the target distribution corresponding to batch data from which the coherent patterns are extracted, the PF recursively solves the sequential data assimilation problem where the posterior distribution evolves at each measurement time through accumulation of the tracer positions. In particular, PF aims to sequentially approximate $\pi(\theta_{0:n},\bx_{0:n}|\bx^o_{0:n})$, where $\bx^o$ denotes the observed tracer positions, $\bx$ is the predicted positions and $\theta$ is the parameter vector. Since $\pi(\theta_{0:n},\bx_{0:n}|\bx^o_{0:n})$ is approximated by the weighted particles $\{(\theta_{0:n}^{(i)},\bx_{0:n}^{(i)}),w_n^{(i)}\}$, the marginalization to obtain the particle distribution $\{(\theta_{n}^{(i)},\bx_{n}^{(i)}),w_n^{(i)}\}$ can be readily done recursively. This particle distribution gives the empirical pdf for $\theta$, $\pi(\theta_{n}|\bx^o_{0:n})$. As for the SMC-ABC method, $n$ is the time index and $i$ the index for each particle. Note that this is an abuse of notation for the sake of convenience, since the time index in the SMC-ABC algorithm is artificial but the time index in PF represents the actual sequence of the data.
\par
In a nutshell, the (bootstrap) PF updates the weight according to
\begin{align}
\label{ISW} w_n^{(i)} \propto& \; w_n^{(i-1)}\,  p( \bx^o_n| \bx_n^{(i)},\theta_n^{(i)}).
\end{align}
As done for the SMC-ABC algorithm, we monitor the Effective Sample Size and resample if $N_{eff}<\delta N$ for some $0<\delta<1$. \rone{We employ an adaptive resampling scheme from \cite{Nakano07}, which
merges particles together in the resampling step. In particular,
instead of resampling $N$ samples from the ensemble, the merging
scheme resamples multiple samples of size $N$ and generates a new
sample of size $N$ using a weighted average of these multiple samples.
This scheme provides a way to gain sample diversity and the weights
are designed to preserve the sample and variance of the original sample. In all of our
experiments, we use 3 samples of size $N$ to generate a new
sample, merged with the weights found in \cite{Nakano07}. }

 For implementation purposes, we will model the likelihood of the observations by
\begin{align}
\label{PFp} p\left( \bx^o_n| \bx_n^{(i)},\theta_n^{(i)}\right) \propto \exp\left(-\frac{1}{2}(\bx_n^o - \bx_n^{(i)})^T \mathbf{R}^{-1}(\bx_n^o - \bx_n^{(i)})\right)\, ,
\end{align}
where $\mathbf{R}$ is the observational error covariance and, on the right of the equation, $\bx_n^{(i)}$ depends on $\theta_n^{(i)}$. As described in the Introduction, we are interested in the highly-informative situation where the drifter locations are measured accurately. As such we take $\mathbf{R} = 10^{-4} \, \mathbf{I}$ in the following experiments.

%%%%%%%%%%%%%%%%%%%%%%%%%%%%%%%%%%%%%%%%%%%%%%%%%%%%%%%%%%%%%%%%%%%%%%%%%%%%%%%%
\section{Case Study: Meandering Jet}
\subsection{Coherent pattern}
\rtwo{Our test model follows the example in~\cite{SamelsonBook}, Chapter 5. We consider a kinematic traveling wave model in the co-moving frame $(x_1,x_2)$ that is deterministically perturbed by an oscillatory disturbance and stochastically perturbed only in the $x_1$-direction:
%\begin{equation}\label{eq:Jet}
\begin{align}
\label{eq:Jetx}
dx_1 &= c - A\sin(Kx_1)\cos(x_2)+\eps l_1\sin(k_1(x_1-c_1t))\cos(l_1x_2)+\sigma dW\\
\label{eq:Jety} dx_2 &= AK\cos(Kx_1)\sin(x_2)+\eps k_1\cos(k_1(x_1-c_1t))\sin(l_1x_2),
\end{align}}
%\end{equation}
where the additive noise is described by a Wiener process $dW$ with variance $\sigma^2$. Here $A>0$ is the amplitude, $c$ is the phase speed of the propagation of the primary wave and $K$ is the wavenumber in the $x_1-$direction. The paramter $c_1$ is the phase speed of the peturbation term in the co-moving frame, $k_1$, $l_1$ are the the $x_1$- and $x_2$- wavenumbers, respectively. We assume the periodic boundary $[0,2\pi)$ in the $x$-direction. For $\sigma=0$, $\eps=0$ and $c$ chosen appropriately, the flow is integrable and consists of two closed circulation regions adjacent to the rigid boundaries $x_2=0$ and $x_2=\pi$ and a jet flowing regime in the $x_1$-direction passing between the two circulation regions. The fluid exchanges under the lobe dynamic due to the oscillatory disturbance of~\eqref{eq:Jetx}--\eqref{eq:Jety} are given in~\cite{SamelsonBook}. \\ In the following we use $\sigma^2=0.01$, $\eps=0.3$, $A=1$, $K=1$, $c=0.5$, $k_1=1$, $l_1=2$, $c_1=\pi$, as given in \cite{SamelsonBook}. \\
\rtwo{For these parameters, the flow around the two circulation regions is dominated on a long time scale by chaotic advection \cite{PierreHumbert91}. A line of drifters placed in one of the gyres and initially spaced $\mathcal{O}(10^{-2})$ apart, that is spaced according to the observation error covariance $\mathbf{R}$, will elongate into a filament that persists to $t\approx 20$. By $t\approx30$ the filament breaks and the drifters appear to be distributed homogenously within the gyre. The flow in the jet between the gyres is not dominated by chaotic advection on these time scales. }

\noindent
\rtwo{We now establish that for this flow there is a coherent pattern, that the coherent pattern can be detected using relatively few tracers, and that the Hellinger distance between simulated and observed coherent patterns is robust to chaos in a sense that the metric employed by the particle filter is not.}

\begin{figure}[htbp]
    \centerline{\includegraphics[scale=.55]{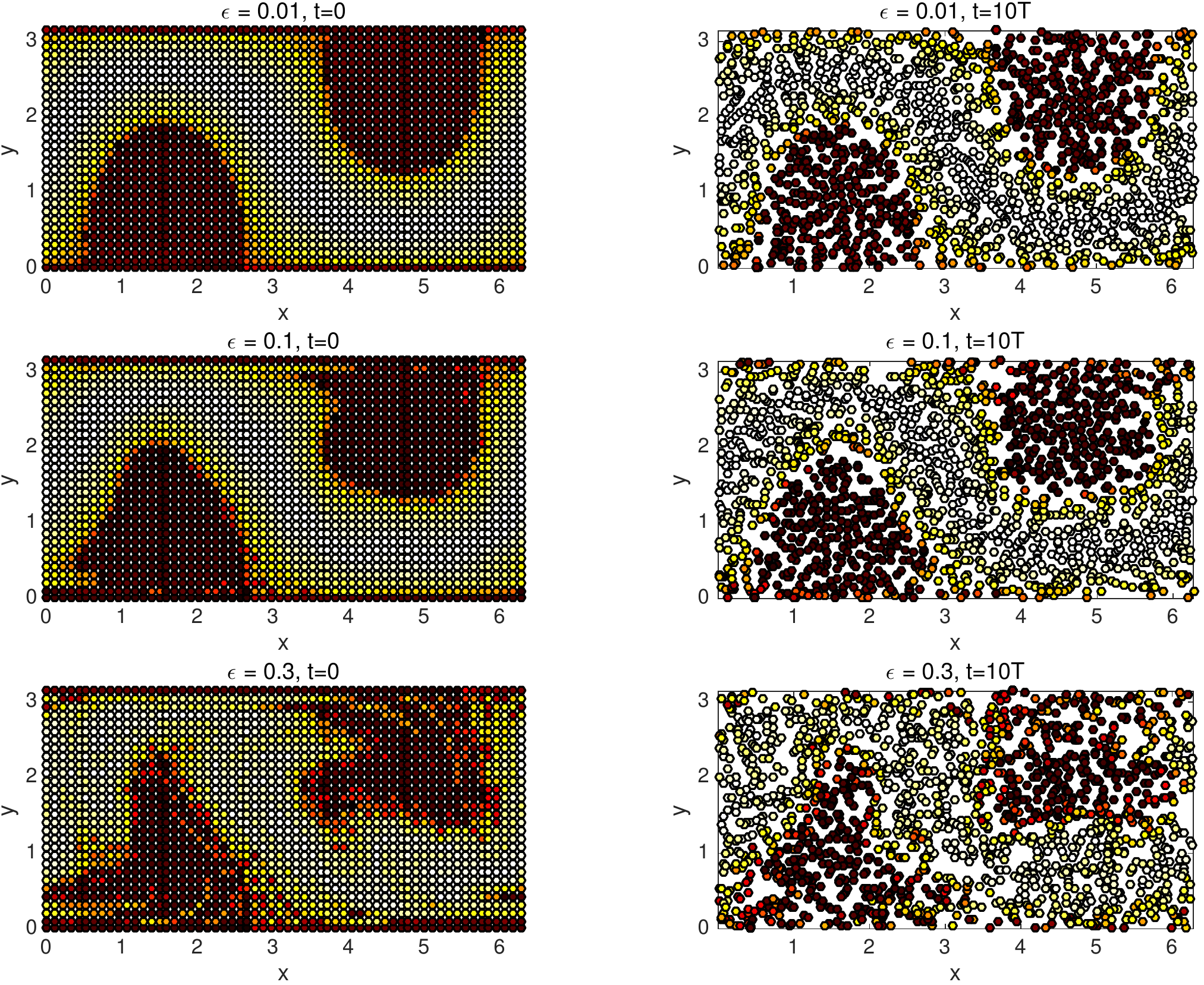}}\caption{Coherent pattern based on the dominant eigenvector of $\bP_{x_1}$ for various values of $\eps$. Left: initial conditions evenly spaced on a grid. Right: flow at time $t=20.$}\label{fig:PCAjet}
\end{figure}

\noindent
 In Figure~\ref{fig:PCAjet}, we show the PCA-based coherent pattern for various values of $\eps$ and for $\sigma=0$.  We initialize tracer positions using $50\times25$ uniform grid points on $(0,2\pi)\times(0,\pi)$ and integrate them to the time $t=10T$ for $T=2$ (the period of the flow under the above parameter values) using the standard Euler-Maruyama method with step size $\delta t=0.01$. The integration time is chosen to be long enough to capture distinct dynamics of the tracers and to inspect the robustness of the coherent pattern. The covariance matrix $\bP_{x_1}$ from~\eqref{eq:Px} is then constructed from the coordinates of the tracer trajectories taken as observations every $10\delta t$ time steps. As shown in Figure~\ref{fig:PCAjet}, although changing its shape with $\eps$, the coherent pattern is very persistent in the sense that those initial tracers labelled in similar colours according to the PCA dominant mode are still tightly grouped together and retain their collective geometrical structure after a long period of time; this gives an intuition of what we mean here by the coherent pattern. Note also that, dynamically speaking, under the small (deterministic) disturbance, the flow is near-integrable and the coherent pattern is determined by the persisting geometry of the circulation regime and jet flowing regime.

\begin{figure}[htbp]
    \centerline{\includegraphics[scale=.425]{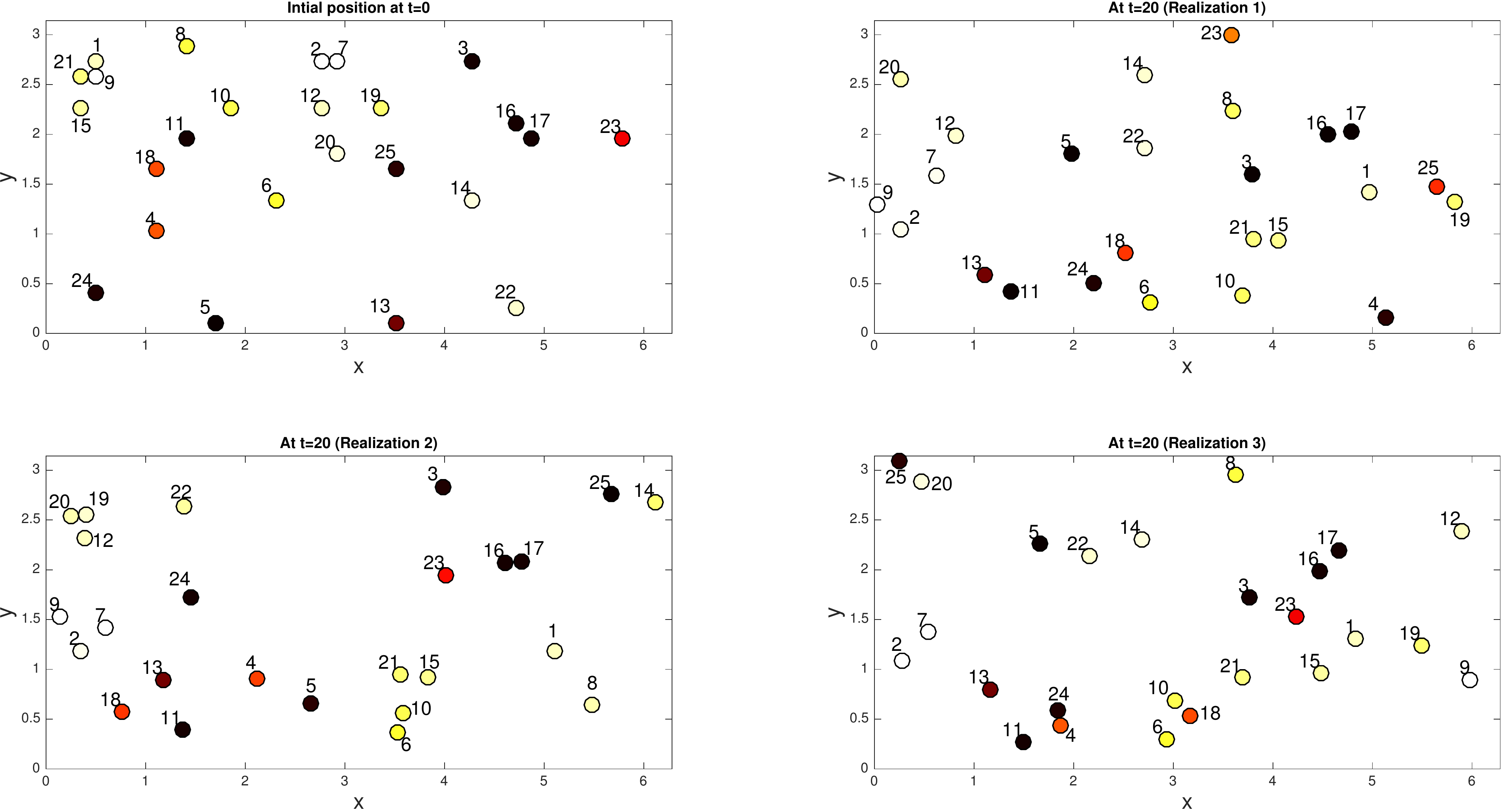}}\caption{\rtwo{The coherent patterns are obtained via PCA for three independent experiments. \emph{Top left}: initial conditions for the other 3 plots, with tracers labelled 1 to 25. \emph{Top right and bottom row}: 3 realizations of the tracer locations at $t=20$, obtained by numerically integrating \eqref{eq:Jetx}--\eqref{eq:Jety}. The realizations are coloured by their PCA dominant mode. We observe that, despite the tracers having very different final locations, the PCA dominant mode tends to assign similar colours to each tracer in all realizations (e.g. tracer 18 is red for all three realizations). } }\label{fig:label_ini}
\end{figure}

\rtwo{We now confirm that the coherent pattern can be detected using relatively few tracers, and that the coherent pattern is in this case robust to the noise and chaotic advection in the dynamical system. We investigate this by limiting the number of tracers to 25, initially randomly placed. We label each tracer and use three realizations of the noise to propagate the tracers to the time $t=10T$. The coherent pattern is again extracted by using the dominant mode of PCA, and the tracers in each of the three realizations are coloured accordingly. We consider PCA to be robust to the chaotic, stochastic dynamics if tracers with the same label have similar colours across each of the 3 realizations. We show the initial conditions and 3 typical realizations in Figure~\ref{fig:label_ini}. }

\begin{figure}[htbp]
\centerline{
    \begin{subfigure}[b]{0.3\textwidth}
    \includegraphics[width=\textwidth]{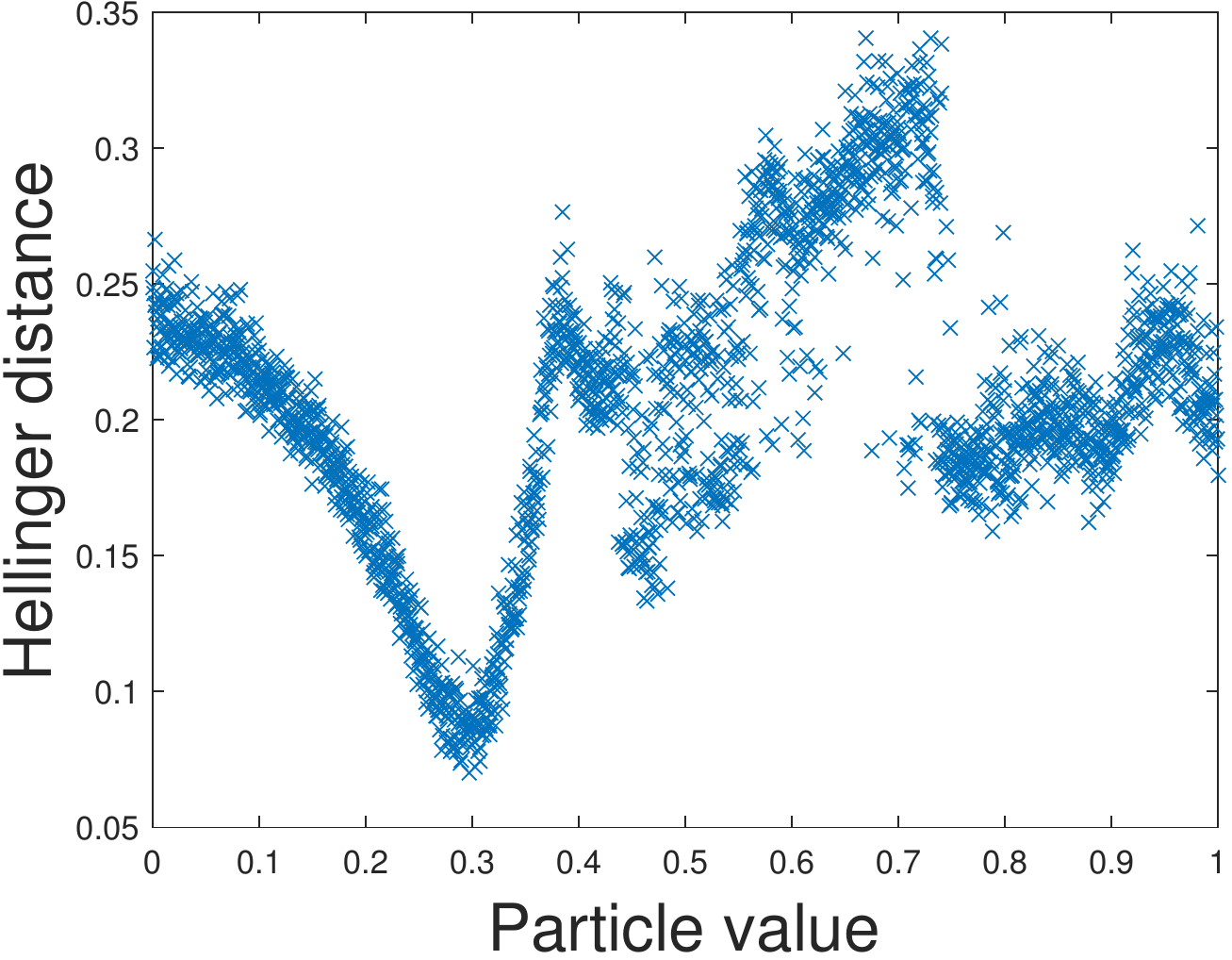}
    \end{subfigure}\hspace{1cm}
    \begin{subfigure}[b]{0.3\textwidth}
    \includegraphics[width=\textwidth]{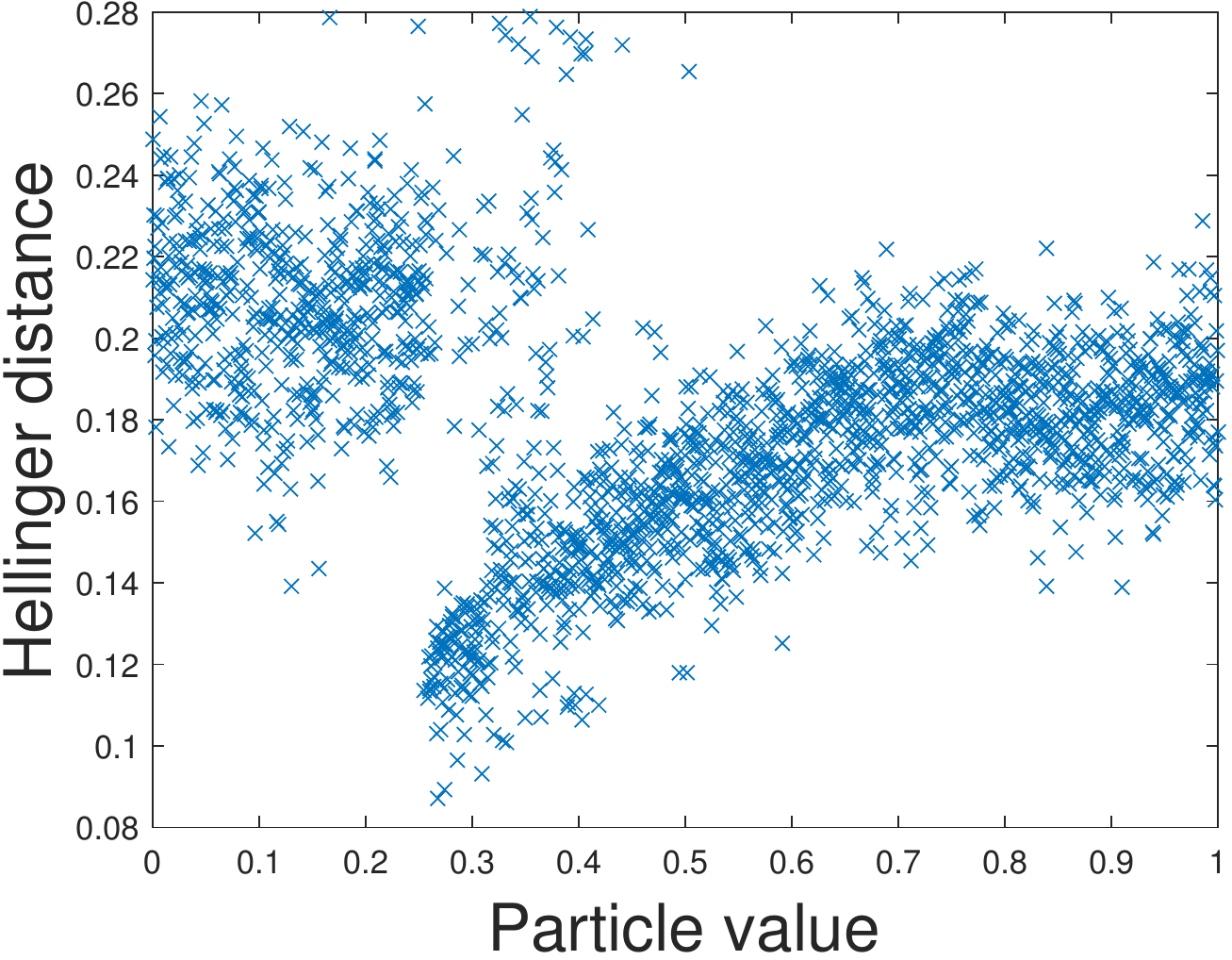}
    \end{subfigure}}
    \centerline{\begin{subfigure}[b]{0.3\textwidth}
    \includegraphics[width=\textwidth]{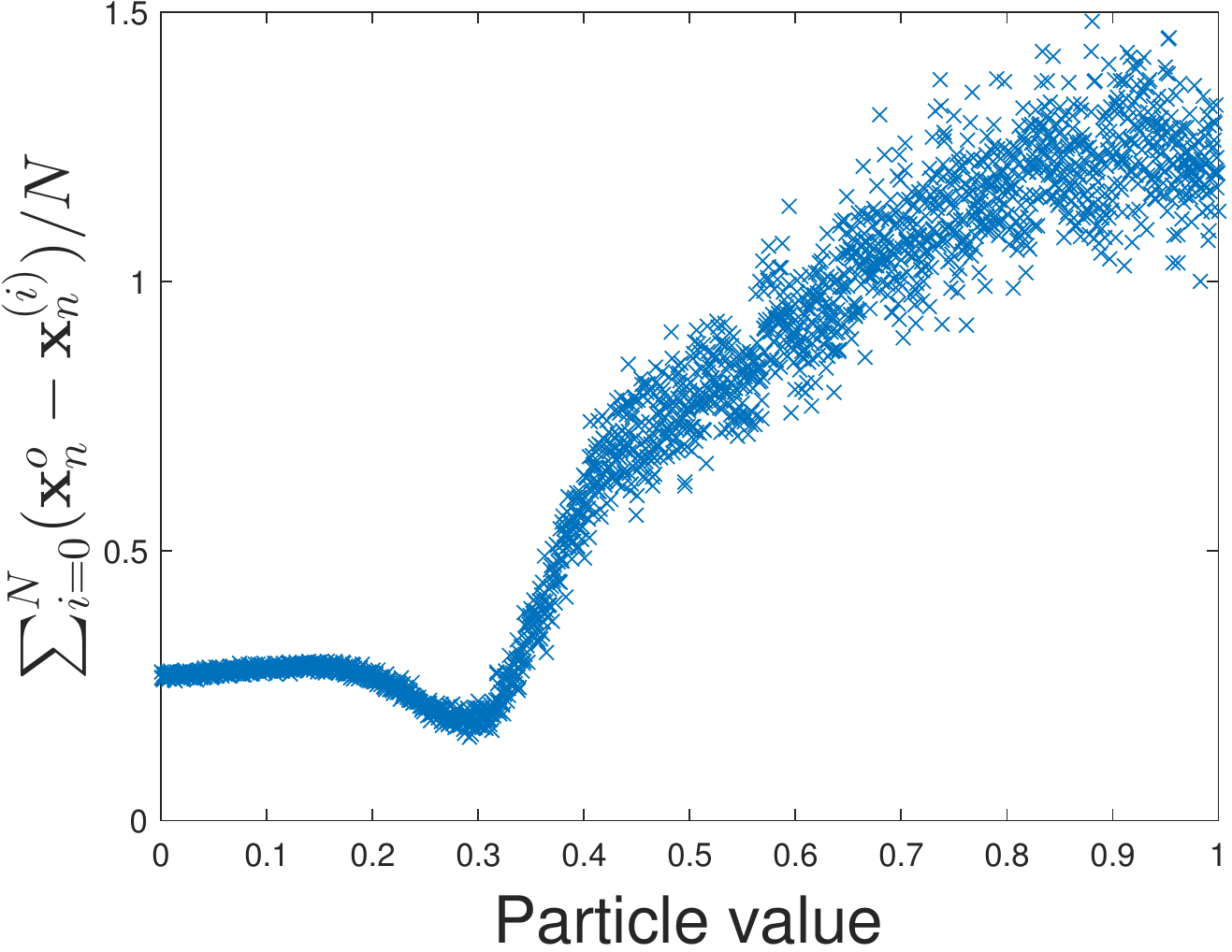}
    \caption{t=20 }
    \end{subfigure}\hspace{1cm}
    \begin{subfigure}[b]{0.3\textwidth}
    \includegraphics[width=\textwidth]{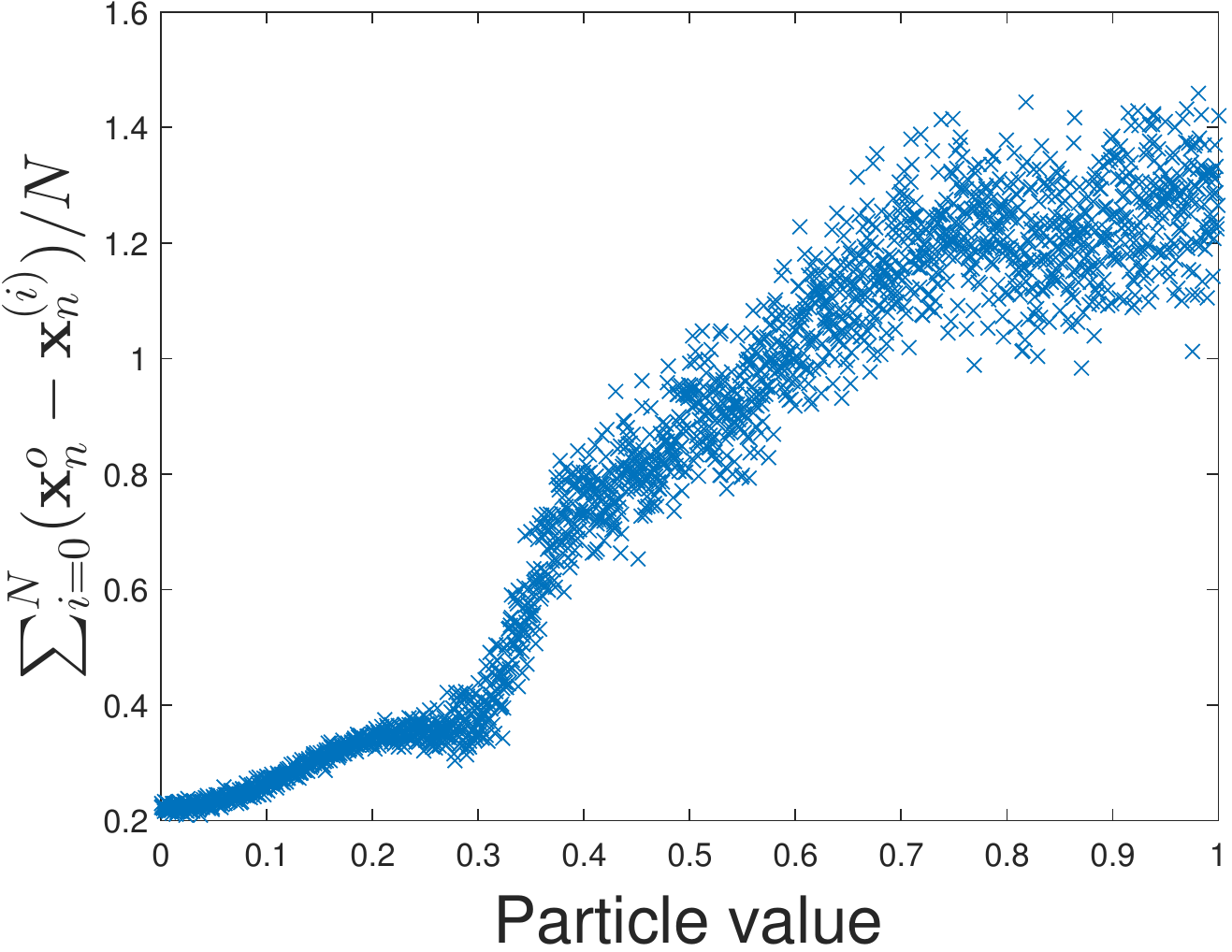}
    \caption{t=30 }
    \end{subfigure}}
    \caption{An experiment was performed in which 50 drifters were numerically integrated by \eqref{eq:Jetx}--\eqref{eq:Jety} using 2000 uniformly spaced values of $\eps$. Each plot shows the particles, i.e. the possible values of $\eps$, plotted against the Hellinger distance (the metric used in SMC-ABC) between simulated and observed patterns, or the distance $\bx_n^o - \bx_n^{(i)}$ between simulated and observed drifters, related to the metric used in PF. For an observation taken at $t=20$, both metrics identify the most likely value of $\eps$ with a minimum at the true value $\eps=0.3$. For an observation taken at $t=30$, only the metric used in SMC-ABC preserves the minimum near the correct location; the metric used in PF would have a minimum at $\eps=0$ corresponding to the unpeturbed flow. }\label{fig:Hellinger}
\end{figure}

%The singular values of $\bP_{x_1}$ for $A=1$ for different values of $\sigma$ are plotted in Figure~\ref{fig:jetspectrum} and it is clear that a large spectral gap occurs between the first and second singular values. This suggests that there is only one significant coherent pattern in this case. The PCA spectrum of the other values of $A$ in Figure~\ref{fig:PCAjet}, not shown here, have a similar trend. 
\rtwo{We now establish the usefulness of the Hellinger distance as a metric to infer model parameters from `observed' and simulated patterns, compared to the direct comparison of the simulated and observed drifter locations used by the particle filter in \eqref{PFp}. }Figure~\ref{fig:Hellinger} compares these two metrics, using a single observation taken at either $t=20$ or $t=30$. We observe that at $t=20$, i.e. before chaotic advection has separated simulated drifters from observed, both the Hellinger distance and the distance between simulated and observed drifters has a clear minimum at the true value $\eps=0.3$. If the observation is instead taken at $t=30$, after chaotic advection has separated the simulated and observed tracer trajectories, then the distance between simulated and observed drifters no longer has a minimum near the correct value of $\eps$, but the Hellinger distance does. We used 50 drifters deployed in a circle of radius 0.1 around the centres $\left(\frac{\pi}{2},1\right)$, $\left(\frac{3\pi}{2},\pi-1\right)$ of each gyre.  
%\begin{figure}[htbp]
%    \centerline{\includegraphics[scale=.5]{spectrum.pdf}}\caption{The spectrum of $\bP_{x_1}$ for $A=1$. }\label{fig:jetspectrum}
%\end{figure}

%%%%%%%%%%%%%%%%%%%%%%%%%%%he spatial pattern%%%%%%%%%%%%%%%%%%%%%%%%%%%%%%%%%%%%%%%%%%%%%%%%%%%%%
\subsection{Data assimilation by SMC-ABC}
In this section we explore the robustness of the SMC-ABC algorithm, measured against the standard particle filter (PF). Following the discussion in Section~\ref{Intro} and the results of Figure~\ref{fig:Hellinger}, we define two `failure states' for the PF. \new{We expect the PF to become degenerate when there are many accurately observed drifters, meaning that one particle will have weight 1 and all other particles will have weight 0. However, we are estimating static parameters; it is quite possible for the (degenerate) particle filter to accurately estimate these parameters with a single particle. We would like to observe a second failure on the part of the PF, that if the drifters are located in regions of chaotic advection in the model, and if the time between observations is sufficient for the initial observation error to completely separate the simulated drifter trajectories from the true trajectories, then the PF may be expected to gain no information over the prior guess for the parameter(s). \\
We focus on estimating the parameters $\eps,\,k_1,\,l_1,\,c_1$ (particularly $\eps,\,k_1$) that characterise the time-dependent peturbation. In order to observe the above failure state in the PF, we will typically initialise the drifters in two rings around the centre of each gyre, and take observations on intervals at least $t=30$. By taking observations on time intervals less than 30, or as we shall see by initialising the drifters uniformly in phase space, the degenerate PF can estimate parameters more accurately than SMC-ABC. It is specifically the robustness of the coherent pattern to chaotic advection that will provide an advantage to SMC-ABC. }
%the spatial pattern of the drifters is robust under noise and chaos (see Figure~\ref{fig:PCAjet}), as is the Hellinger distance between the observed and simulated coherent patterns as estimated by PCA (see Figure~\ref{fig:Hellinger}); furthermore the estimate of the coherent pattern is expected to improve as the number of drifters increases. Therefore the SMC-ABC method should still work with many drifters \rtwo{and a chaotic dynamical system}.\\

%\noindent
\subsubsection{Experimental setup}\label{setup}
\new{In the following experiments, we use a final time of $t=30$. The PF and SMC-ABC schemes use a model time step of $0.1$, and a single observation is taken at the final time. The observations are synthetically generated from a model integration using a time step of $0.01$.} \\Our observations are of the trajectories of $50$ tracers; the initial positions of these tracers are typically circles of radius 0.1 around the centres of each gyre. We construct the ``truth" by taking a realization produced by the above parameter values.\\

\noindent
\new{We are interested in comparing the performance of PF and SMC-ABC at the same computational cost. Accordingly, we modify SMC-ABC as follows: we choose the number of particles $N$ in SMC-ABC to be one tenth the number of particles used in the PF. Furthermore, we restrict the SMC-ABC algorithm to at most ten steps in the artificial time $n$; that is if $n=10$ in Algorithm 1 the While loop is halted and we return the final particle distribution $\{\theta^{(i)}_{10}\}$. With this modification the computational cost of SMC-ABC is controlled to be roughly equivalent to, or less than, the cost of PF.} \\

In the following experiments, the transition kernel $q_n(\theta,\theta^\ast)$ is given by the random walk
\begin{equation}\label{eq:randomwalkpara}
\theta^\ast = \theta + \eta,\qquad\qquad\eta\sim\mathcal{N}(0,0.01).
\end{equation}
Therefore, the MH ratio~\eqref{eq:MHratio} is either 0 or 1; the weight of the $i-$th particle at the $n-$th step is 1 if it has survived all $n$ rejection steps, or if it has been resampled and then survived the intervening rejection steps, and it is 0 otherwise.

%We label the parameters to be estimated by $\bq$. In the one-parameter experiments, we assume that we wish to estimate only $\bq=\eps$ by assimilating the coherent spatial pattern revealed by the tracer trajectories via PCA. We recall that although the likelihood probability density of the tracer observation~\eqref{PFp} is completely known in this case, the likelihood of the coherent patterns is unknown. Therefore, we will employ SMC-ABC to estimate $\bq$. \\

%%We will report the absolute error for experiments with the PF and SMC-ABC; this is the distance between the sample average $\sum_{i=1}^N \theta_n^{(i)} w_n^{(i)}$ and the true parameter(s) $\bq$. We will also typically report the average performance of, and variance in, multiple repetitions of each numerical method.

\subsubsection{One parameter experiments}
\rtwo{We perform experiments to estimate the time-dependent peturbation parameter $\eps$, assuming that the prior density $\pi(\eps)$ is a uniform distribution on the interval $[0,1]$.}\\

\begin{figure}[htbp]
  \centerline{  \includegraphics[width=0.5\textwidth]{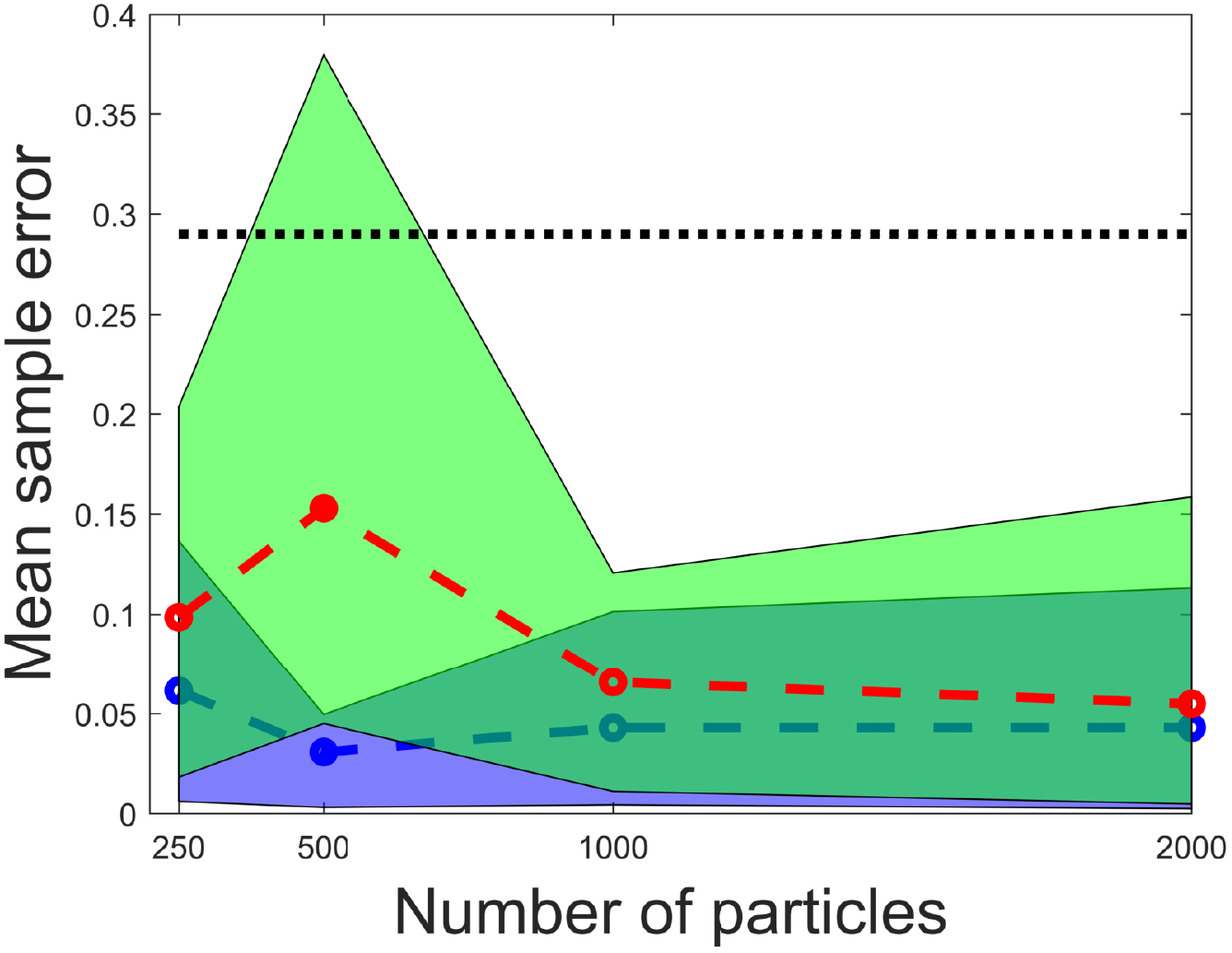}\includegraphics[width=0.5\textwidth]{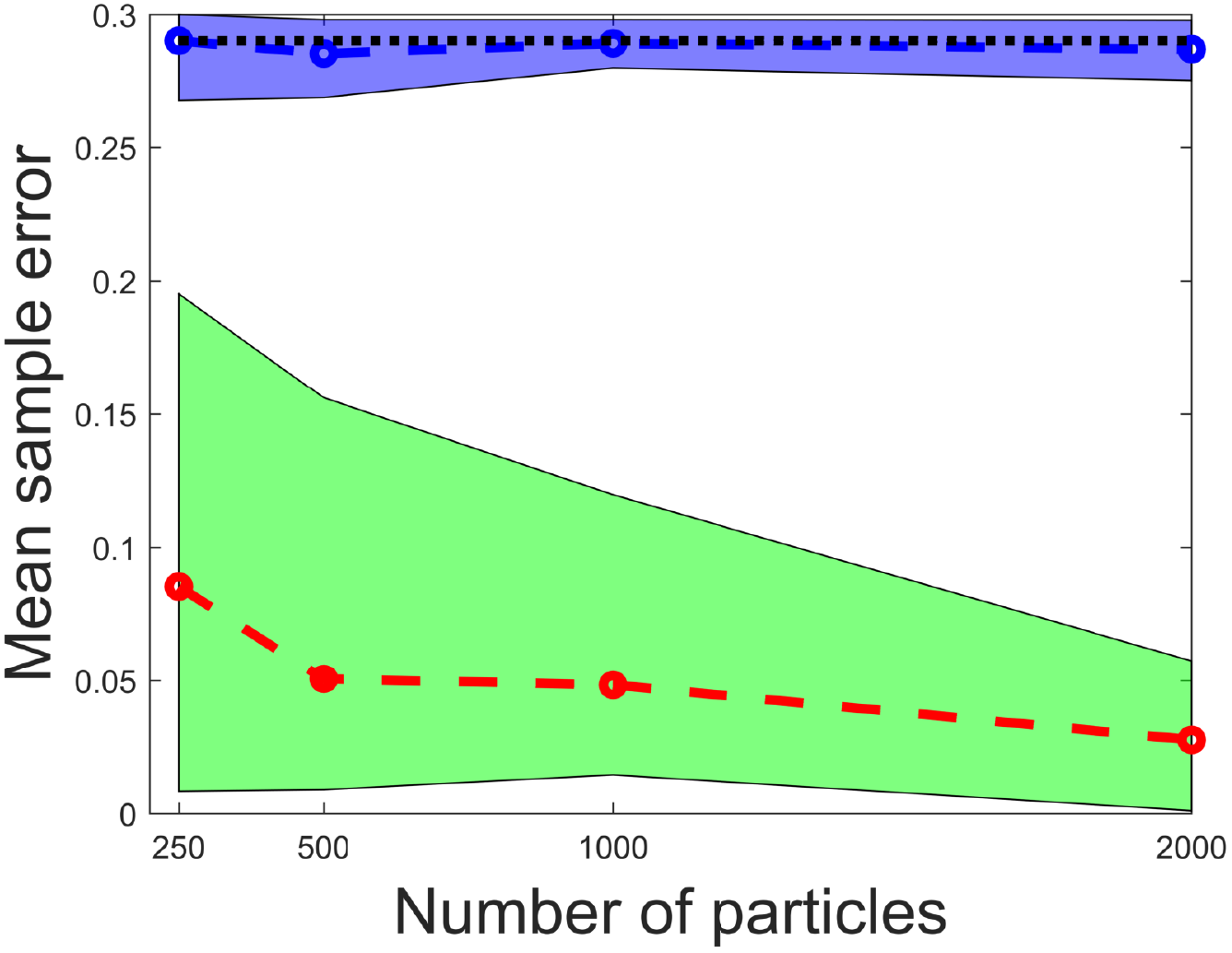}}
    \caption{Error in estimating the peturbation strength $\eps$ in the PF {\textbf{\color{Cerulean}(blue)}} and SMC-ABC {\textbf{\color{ForestGreen}(green)}}. Circles and dashed lines show the mean error of 20 repetitions from each numerical method, while the coloured patches show the $10\%$ and $90\%$ percentiles. The black dotted line shows the mean error of the initial distribution of the particles. \emph{Left: }drifters are initially deployed uniformly, and the Particle Filter (while always degenerate) can accurately estimate $\eps$. \emph{Right: }drifters are initially deployed within the gyres, their trajectories are dominated by chaotic advection, and the Particle Filter gains no information on $\eps$ from the drifter locations. In both cases SMC-ABC can infer $\eps$ from the coherent pattern created by the drifters.}\label{fig:varyN}
\end{figure}

\rtwo{
We investigate how the absolute error $\left|\sum_{i=1}^N \theta_n^{(i)} w_n^{(i)}-\eps\right|$ scales with the number of particles $N$, when all other parameters are kept fixed. We present results for PF with $N = \{250, 500, 1000, 2000\}$ particles, and SMC-ABC with $N =\{25, 50,100,200\}$ particles. As discussed in Section~\ref{setup}, under this set-up SMC will have lesser or equal computational cost to PF, and we overlay the results accordingly. We suppose that the true parameter is $\eps=0.3$, and present results for two situations, one in which the drifters are initially distributed uniformly and a second in which the drifters are initially distributed in a circle of radius 0.1 around the centres $\left(\frac{\pi}{2},1\right)$, $\left(\frac{3\pi}{2},\pi-1\right)$ of each gyre. These drifters will all be contained within the gyres (see Figure~\ref{fig:PCAjet}). As discussed previously, in the first situation some drifters will be in nonchaotic or weakly chaotic regions of the flow, and the PF should gain information; in the second situation all trajectories will be dominated by chaotic advection and the PF should be totally uninformative. Figure~\ref{fig:varyN} clearly shows the expected behaviour from PF, highlighting that SMC-ABC is unaffected by the chaotic flow. We mention that each run of the PF is degenerate (having one particle with weight 1).}\\
%We choose $\alpha=0.2$, and the scale separation parameter $eps=10^{-9}$. We use $M=40$ microsteps with microstep size $dt = 0.4eps$, while the number of iterations $n$ vary from $20$ to $10^5$ to keep $T=1$ fixed for all values of $Dt$. Initial conditions are chosen to lie on the approximate slow centre manifold with $y^0 = 1$, $x^0 = \sin^2(1)$. The Lipschitz constants are $L_g=1.1$ and $L_h = 1$, the bound on the vector field of the slow dynamics is $C_g=2$, and the maximal derivatives of the reduced slow dynamics are $C_2^*=4$ and $C_4^* = 8$.\\

\subsubsection{Multiple parameter experiments}
In this section we use SMC-ABC and PF to simultaneously estimate the perturbation parameter $\eps$ and the amplitude and frequency parameter $k_1$ in \eqref{eq:Jetx}--\eqref{eq:Jety}. We label the set of parameters to be estimated by $\bq=(\eps,k_1)$, with true values $\epsilon=0.3$ and $k_1=1$. We take the prior density $\pi(\bq)$ to be a uniform distribution on the interval $[0,1]\times[0.2,1.2]$.\\

\par
\begin{figure}[h!tbp]
    {\includegraphics[width=0.36\textwidth]{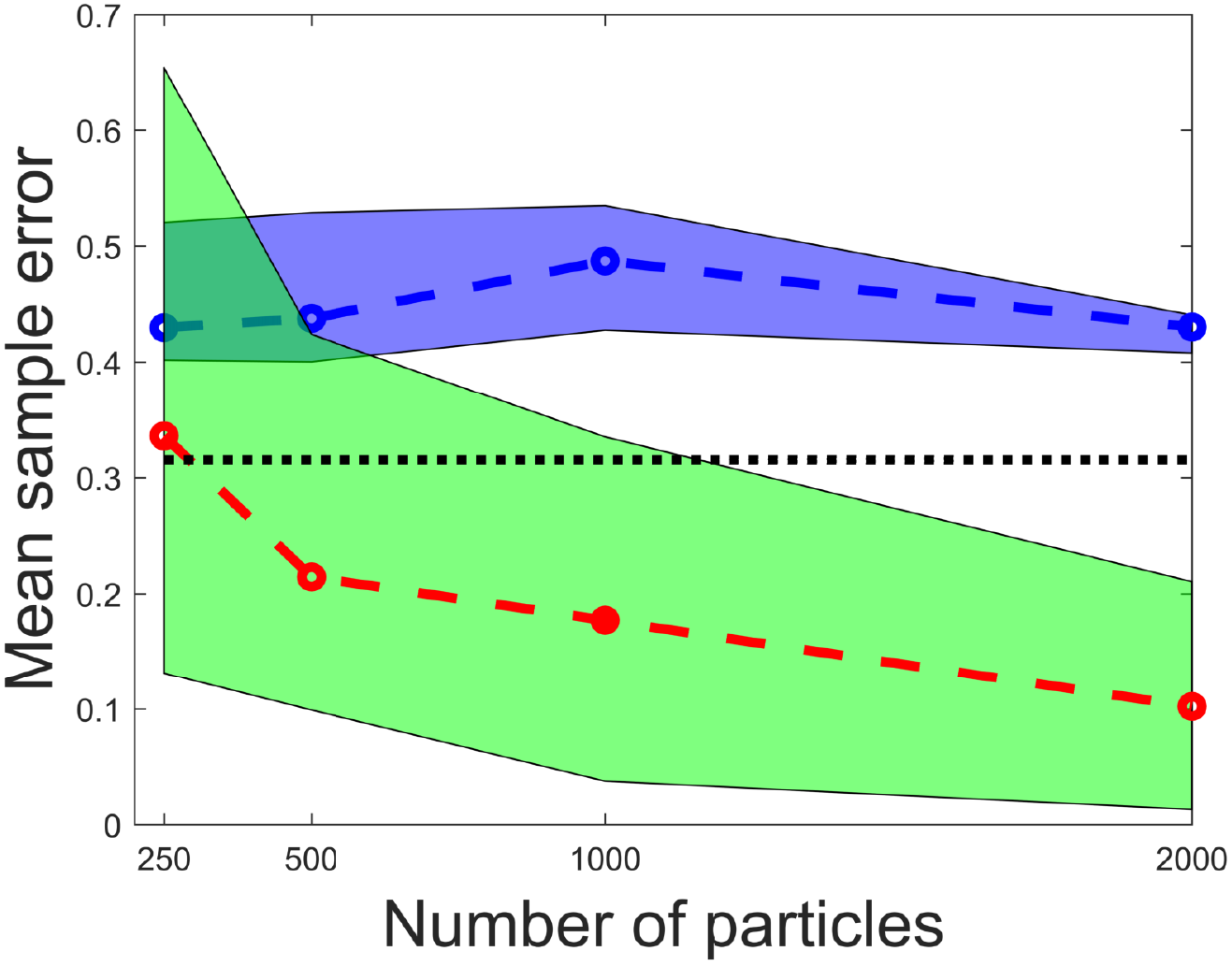}\hspace{1cm}\includegraphics[width=0.45\textwidth]{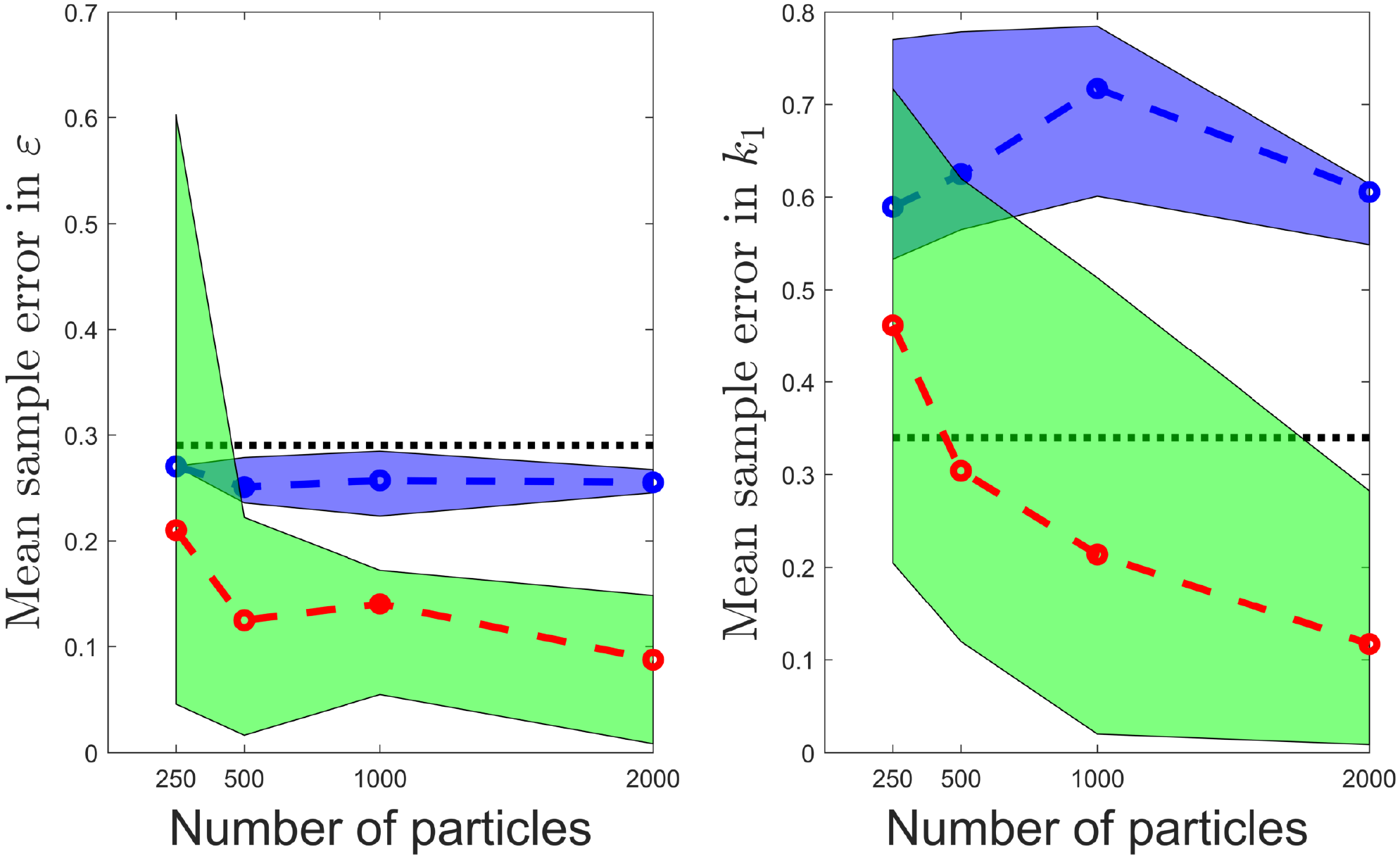}}
\caption{\emph{Left:} Mean error in estimating parameters in the PF {\textbf{\color{Cerulean}(blue)}} and SMC-ABC {\textbf{\color{ForestGreen}(green)}}. Circles and dashed lines show the mean error of 20 repetitions from each numerical method, while the coloured patches show the $10\%$ and $90\%$ percentiles. The black dotted line shows the mean error of the initial distribution of the particles. \emph{Left: }Mean error in estimating both $\eps$ and $k_1$. \emph{Right}: Error in estimating the individual parameters. In both cases the PF performs poorly and SMC-ABC, while having a large variance, improves on the prior guess and can accurately estimate both parameters.}
    \label{fig:multi}
\end{figure}

We investigate how the absolute error $\left|\sum_{i=1}^N \theta_n^{(i)} w_n^{(i)}-\bq\right|$ scales with the number of particles $N$, when all other parameters are kept fixed. The drifters are again initialised inside the boundaries of each gyre, as done previously. We present results for 20 runs each of a PF with $N = \{250, 500, 1000, 2000\}$ particles, and SMC-ABC with $N =\{25, 50,100,200\}$ particles. Again we overlay the results in view of their roughly equivalent computational cost. Figure~\ref{fig:multi} confirms that SMC-ABC can accurately estimate multiple parameters, while again the PF fails to improve over the prior guess. We remark that the broad spread of the results for SMC-ABC compared to PF is a consequence of the adjustment to the same computational cost; SMC-ABC is (also) a sequential monte carlo method employing one tenth the number of particles as PF, and consequently outputs particles with relatively high variance.

%We now produce a summary Table for the error statistics of the PF and SMC-ABC as done for the one parameter experiment. As before we vary the number of particles, the number of drifters, and the magnitude of the true perturbation parameter $\eps$. The SMC-ABC method successfully jointly estimates $A$ and $\eps$ for all of the chaotic flows. The results are in Table~\ref{tab:3p}.
%
%\begin{table}
%\begin{center}
%\begin{tabular}{c  c c c c c c}
%$\eps$  & No. of particles  & No. of drifters & PF error & SMC-ABC error  \\
%\hline
%     0   &   500   &   20   &     0.22 &   0.18
%\\
%    0  &  500    & 50      &   0.31     & 0.06 
%\\
%0   &  2000  &   20     &   0.24   &   0.39
%\\
%    0   &   2000  &  50    &   0.29   &   0.04
%\\
%     0.3 &  500   &   20    &    0.24    & 0.06 
%\\
%   0.3 & 500   &  50    &     0.20      & 0.05
%\\
%     0.3  &   2000   &  20   &    0.24     & 0.07
%\\
%    0.3  &  2000   & 50 &       0.31        & 0.02
%\end{tabular}
%\end{center}
%\caption{Parameters and accuracy statistics for 20 runs of the PF and SMC-ABC schemes, for a range of parameters. The error statistic presented is the mean error for both parameters.}
%\label{tab:3p}
%\end{table}

\section{Discussion}
This paper presents, to our best knowledge, the first attempt to develop an approach to assimilate the Lagrangian coherent structure for parameter estimation. In contrast to previous research for LaDA, the new approach does not directly use the error statistics of observed positions of the tracers in order to make a parameter inference but assimilates the coherent structures as observations instead. Therefore, our approach assumes the existence of coherent structures and requires a tool that can extract that structure from the trajectory data. The structure is extracted via a machine learning method, which is PCA in the current work, and the assimilation algorithm is based on the Approximation Bayesian computation (ABC), which allows a bayesian inference to be performed without computing the typically unavailable likelihood function of the coherent structure. However, a distance function must be appropriately designed to measure the discrepancy between the observed coherent structure and the coherent pattern predicted by the model simulation based on various parameter values. We use the Hellinger distance and empirically demonstrate its robustness in the situation where the underlying flow exhibits chaotic advection and is stochastically perturbed.
\par
\rtwo{Our numerical experiments demonstrate that this new approach is remarkably superior to the trajectory-based Lagrangian DA (employing the particle filter) in the situation where the number of tracers is large and the drifter trajectories are dominated by chaotic advection. In order to achieve this situation, we initialised drifters in the chaotic regions of a peturbed jet flow.} The comparison uses the average mean-square error to measure the accuracy of the estimate. The improved accuracy of the ABC approach is attributed in large part to the following facts: (1) the Hellinger distances for the coherent structure is usually less sensitive to the stochastic perturbation than the likelihood function of tracer trajectories, which is assumed to be highly informative (i.e. to have small variance). (2) Even in the chaotic flow, the large-scale coherent structure is still geometrically ``regular" while the tracer trajectories may experience high dispersiveness due to the stretching and folding process.
\par
\rone{We now briefly discuss convergence results for the SMC-ABC algorithm. On the SMC side,~\cite{DelMoralDoucetJasra12} discusses how to obtain convergence results for the stochastically resampled SMC scheme by connecting it to deterministic resampling schemes (convergence results for which are obtained via the Central Limit Theorem and are given in, for instance, \cite{DoucetEtAl01}). It may be more difficult to establish convergence for the ABC component, which would typically entail confirming that the coherent pattern (as approximated by some numerical method) is a summary statistic for the tracer statistics, in the sense that the likelihood function for the coherent pattern is the same as the likelihood function for the tracer trajectories; see for instance the Discussion in \cite{MarinEtAl12}.} \new{We comment that for the flow considered in this paper, the coherent pattern is more than just a summary statistic. In other words, the likelihood function given by the tracer trajectories is different to the likelihood function given by the coherent pattern. }
\par
In future work we will examine our approach in the case of transitory coherent sets, which exist for a finite time before losing coherency due to a change of parameters. Detecting the parameter change is of interest in many applications and it will require developing a new algorithm that exploits the coherent structure to track the change.

\section{Acknowledgements}
JM is grateful for the support of the MURI grant A100752, ONR grant N00014-15-1-2112, and NSF grant DMS-1312906. NS acknowledges the support of the Department of Mathematics at Univeristy of Surrey.

\bibliographystyle{plain}
\bibliography{LADAbib}

\end{document}